# On The Liniar Time Complexity of Finite Languages.

Mircea Alexandru Popescu Moscu
Bucharest, Romania
mircalex@yahoo.com

**Astract**

The present paper presents and proves a proposition concerning the time complexity of finite languages. It is shown herein, that for any finite language (a language for which the set of words composing it is finite) there is a Turing machine that computes the language in such a way that for any input of length k the machine stops in, at most, k + 1 steps.

**Proposition 1:** For any finite language L over an alphabet $\Sigma$ there exists a Turing machine M that accepts the language L; in addition, the machine M has the following property: for any natural number k and for any word w of length k over $\Sigma$ the machine M stops in, at most, k + 1 steps.

**Proof:**

We confine ourselves only to languages over the alphabet {0,1} since the generality of the proof is not affected.

Let L be a finite language over $\Sigma = \{0,1\}$
We are going to construct now the Turing machine M that accepts this language and has, in addition, the property found in our statement.

Since the language L is finite, there exists a number n such that any word in the language L has a length lower or equal to n.

In fact $n = \max(\{|w| \mid w \in L\})$ where $|w|$ denotes the length of w.

The machine is defined like this:
$M = (\Sigma, \Gamma, Q, \delta)$.
$\Sigma = \{0,1\}$
$\Gamma = \Sigma \cup \{b\}$
$Q = \{q_{start}, q_{accept}, q_{reject}, q_0, q_2, ..., q_{2^n-1}\}$
$\delta: (Q - \{q_{start}, q_{accept}\}) \times \Gamma \to Q \times \Gamma \times \{-1, 1\}$

if $\lambda \in L$ (the word with no letters, nil word)
    $\delta(q_{start}, b) = (q_{accept}, b, 1)$
else
    $\delta(q_{start}, b) = (q_{reject}, b, 1)$

We consider that the words over {0,1} are natural numbers represented in base 2. We consider that these representations of the numbers are arranged backwards on the tape.

$\delta(q_{start}, 0) = (q_0, b, 1)$
$\delta(q_{start}, 1) = (q_1, b, 1)$

for any k from 0 to $2^n-1$, let $(k_i)$ with i from 0 to $\lceil \log_2(k) \rceil - 1$ be the representation of k in base 2.

for any k from 0 to $2^{n-1}-1$ and for any $x \in \{0,1\}$ $\delta(q_k, x) = (q_p, b, 1)$, where p has the representation $(p_i)$ with i from 0 to $\lceil \log_2(k) \rceil$ and $p_i = k_i$ for all i from 0 to $\lceil \log_2(k) \rceil - 1$ and $q_h = x$, where $h = \lceil \log_2(k) \rceil$

for any k from $2^{n-1}$ to $2^n-1$ and for any $x \in \{0,1\}$ $\delta(q_k, x) = (q_{reject}, b, 1)$

for any k from 0 to $2^n-1$, if $w \in L$, where $w = (p_i)$ and i goes from 0 to $z = \lceil \log_2(k) \rceil - 1$ and $p_i = k_{z-i}$, then $\delta(q_k, b) = (q_{accept}, b, 1)$, else $\delta(q_k, b) = (q_{reject}, b, 1)$

It is obvious that the machine M accepts the language L and also that any computation requires only i+1 steps to complete, i being the length of the input word (not counting the blank symbol).

q.e.d

**Conclusion**: We know now that for anything we want to compute there exists a "liniar machine" doing it. But this has no importance since we cannot, in general, construct such a machine; and even if we did the costs would have been to big.

**Reference:**
You can refer to my paper at arxiv for a slightly different version of this result: http://arxiv.org/pdf/cs.CC/0411033